\documentstyle[11pt,aasms4]{article}
\def\etal{{\it et al.\ }}

\def\putplot#1#2#3#4#5#6#7{\begin{centering} \leavevmode
\vbox to#2{\rule{0pt}{#2}}
\includegraphics{#1}

\end{centering}}

\lefthead{Brosch \& Hoffman}
\righthead{IN Comae and LoTr 5}

\begin{document}
\title{The shape of the LoTr 5 planetary nebula}
\author{Noah Brosch\altaffilmark{1}}

\affil{Space Telescope Science Institute \\ 3700 San Martin Drive \\
 Baltimore MD 21218,
U.S.A.}

\and  

\author{Yehuda Hoffman\altaffilmark{2}}

\affil{Kapteyn Astronomical Institute \\
 University of Groningen \\ P.O.Box 800, 9700 AV Groningen \\ The Netherlands}

\altaffiltext{1}{On sabbatical leave from the Wise Observatory and 
the School of Physics and Astronomy,
Raymond and Beverly Sackler Faculty of Exact Sciences,
Tel Aviv University, Tel Aviv 69978, Israel}

\altaffiltext{2}{On sabbatical leave from the Racah Institute of Physics, 
The Hebrew University, Givat Ram, Jerusalem 91904, Israel}


\begin{abstract}
We observed the large and faint planetary nebula (PN) around IN Com
in H$\alpha$ and [NII] light
with a coronagraphic CCD device on the Wise Observatory reflector
blocking the light from the central star. Our goal was to provide
a second image of the object with which to confirm the features seen in the
only published photograph from the paper reporting the discovery of this
object. The nebula is extremely faint,
but a combination of images totalling $\sim$one and a half
hours of exposure shows it fairly well. 
A novel image processing algorithm has been applied 
to the noisy image in order to reveal faint extended details of the images. 
The algorithm is based on a non-linear self-adaptive filter applied to 
the wavelet transform of the image. The nebula is not round or elliptical,
but shows a two-lobed and possibly three-lobed morphology, as well as a
peculiar hole-like feature East of the central star. There
is definite East-West and slightly less definite North-South asymmetry.

\end{abstract}


\section{Introduction}
The planetary nebula (PN) LoTr 5=339.9+88.4, around the star IN Com, was 
discovered by Longmore \& Tritton (1980, LT80). It has the distinction of 
being the highest galactic latitude PN at b$\approx88^{\circ}$ and is 
very faint. Although many publications analyzed the
central star, the only image of the PN appears in the discovery paper 
and is from a  IIIaJ+GG395 plate from the UK Schmidt  ESO/SRC Southern
sky survey; the PN is not visible on the Palomar Sky Survey. 
The image shown by LT80 shows some deviations from roundness as
well as some regions within the PN envelope with fainter
surface brightness. Bond \& Livio (1990) 
count LoTr 5 among the elliptical morphology PNs
and explain this as the influence of an extreme-to-moderate density contrast 
produced by binary ejection, followed by shaping by a fast wind. 
Tweedy \& Kwitter (1996) classify it as a point-symmetric PN and conclude
that it shows no interaction with the ISM.  

LoTr 5 is peculiar  because its central star is a possible triple system.
We were drawn to it following a study of UV sources in the direction of Coma,
observed by the FAUST telescope (Brosch \etal 1998), where IN Com  
was detected as a 1650\AA\, UV source. A literature search revealed that the
star is classified optically as G5III, but IUE spectra show an upward turn  
for $\lambda\leq$2300\AA\, in the spectral energy distribution, to the shortest UV 
wavelengths observed (Feibelman \& Kaler 1983), an indication  
of a very hot body in the system. Various photometric and spectroscopic periods
have been reported for IN Com, ranging from 1$^d$.2 to 5$^d$.9 and
even 2,000$^d$ (Schnell \& Purgathofer 1983;
 Malasan \etal 1991; Jasniewicz \etal 1987 and 1994; Strassmeier \etal 1997a and 1997b).
 
IN Com is an X-ray source (Apparao \etal 1992; Kreysing \etal 1992). The 
emission is
soft and corresponds to a body at 1-2 10$^5$ K. The system is also   
an EUVE source  in the 58--174\AA\, band (Fruscione \etal 1995).
This is presumably one of the hottest known stars (T$_{eff}\geq$120,000K;
Feibelman \& Kaler 1983), but may not be the only X-ray source in the system.

The distance estimates to LoTr 5 range from 80 pc (Buonatiro 1993) to 6.3 
kpc (recalculated by Jasniewicz \etal 1996). The Hipparcos Catalog lists for IN Com 
a trigonometric parallax consistent 
with zero (0.83$\pm$1.17 mas/yr), locating the object at a distance larger than 
$\sim$0.2 kpc (3$\sigma$). A 100 pc distance to IN Com would translate its
Hipparcos-measured proper motion into a transversal velocity of
the same magnitude as its radial velocity (--9 km s$^{-1}$ in Jasniewicz \etal 1994;
--9.6$\pm$0.4 km s$^{-1}$ in Strassmeier \etal 1997).

IN Com has been reported to be a triple star (Malasan \etal 1991), consisting 
of a sdO star
(M$_*\approx1.1M_{\odot}$) orbiting with a $\sim$2000$^d$ period a close binary 
system made of a G5III star and a low-mass star. Jasniewicz \etal
(1996) argued that the close binary consists of the 1.1 M$_{\odot}$ G5III star and
the sdO star (M$_*\approx0.6M_{\odot}$, typical of a sub-dwarf), and did not rule
out an additional period of a few years. The close binary is considered
pre-cataclysmic by Cherepashchuk \etal (1996). An argument that the sdO and the G5III
stars evolved in close neighbourhood is the observation that
Ba, Sr, and Y are overabundant in the giant star's atmosphere (Th\'{e}venin
\& Jasniewicz 1997), possibly the result of s-process elements from the
AGB phase of the present-day sdO star polluting the G star's atmosphere through 
a stellar wind.  Strassmeier \etal (1997) did not find radial
velocity changes in the G5III star and concluded that it must
be the outer distant component of a triple system.

The PN is an evolution product, as is the sdO star, and the shape of the nebula may 
provide a clue to the nature of the central stars of this system. If the sdO, which
is the planetary nebula nucleus (PNN), is very 
distant from the G5III star, as argued by Malasan 
\etal (1991), its evolution should have happened almost in isolation and one 
may expect the PN to be symmetrical. If, on the other hand, the interpretation 
by Jasniewicz \etal (1996) is correct, one should expect the PN to show some
asymmetry. A detailed morphological study of the nebula requires first a
confirmation of the general features seen in the broad-band image published
by LT80, and even better, an image obtained in an emission line to
disentangle continuum emission from the nebula proper. We present here a new 
image of LoTr 5 obtained in H$\alpha$, which contains morphological evidence 
to support the second possibility, namely of the PNN being one component of
a close binary.

\section{Observations}

We  observed IN Com with the Wise Observatory (WiseObs) 40" reflector 
in order to obtain a new image of the nebula. Preliminary attempts to image 
directly the nebula with a 50\AA\, FWHM H$\alpha$ filter whose transmission profile
is centered on 6562\AA\, were unsuccessful.
We attribute this to light scattered within the telescope from the  G5
central star (V=8.95$\pm$0.02, B--V=0.835$\pm$0.004; Hipparcos Catalog), 
which is much brighter 
than the  nebulosity and overloads the light in the image.

WiseObs operates a small coronagraph at the f/13.5 focus of the 40" reflector, 
with which the Shoemaker-Levi 9 collision with
Jupiter, and later the passage of the Earth through the Saturn ring plane, were
observed. The coronagraph is a re-imaging device using a tilted parabolic
mirror as collimator, an occulting finger in the telescope focal plane, 
and an apodizing mask at the location of the pupil. The light is re-imaged 
onto a Tektronix 1024$\times$1024 CCD through a filter mounted before the camera
lens. The plate scale with the coronagraph is 0".88 pixel$^{-1}$
and the field defined by the device's entrance aperture is 630" in diameter.
The projected diameter of the occulting finger is $\sim$9".2 and the finger is
mounted very close to a North-South position in the field of the instrument.

The use of the coronagraph allows blocking out of light from the bright
G5III star (and from a southern companion, $\sim$3 mag fainter),
reducing significantly the amount of parasitic light into the system. The 
disadvantage is that
the instrument, being very simple, has no guiding option. The exposures analyzed 
here are unguided and the stellar images are slightly trailed. The 
trailing was kept small by using short exposures and recentering, and by observing 
close to meridian passage of the object.
 
We  obtained six images H$\alpha$+[NII] of IN Com and its nebula, 
ranging from 300 sec to 1200 sec exposure and contains mostly H$\alpha$ light
(the relative transmission of the filter at [NII] is about half or less
than at the H$\alpha$ line). In total, the images account for 5200 
sec of exposure. The CCD frames were debiased, flat-fielded against 
twilight sky flats,
 zapped to eliminate cosmic rays, registered using a number of stars in
the field, and added together. The resultant image is presented in Figure 1
and is called here ``the H$\alpha$ image'', with the caveat that some
[NII] emission may also be present in the data. The image
 shows a circular area, corresponding to the
illuminated part of the CCD, inscribed in a square which is the entire CCD frame.
A straight feature runs to the right on the image; this is the shadow of
the occulting finger with some light diffracted around it and reaching the detector.
The stellar images in the final image are trailed  $\sim$10" in the E-W direction,
and their FWHM in the N-S direction is $\sim$7". This does not prevent 
the analysis of much larger asymmetries in the image of the PN.

\section{Results}

The image in Fig. 1 shows a roundish nebulosity around the central star of LoTr 5 (IN Com), 
and is very similar to that reproduced in LT80. The central star of the PN is mostly 
occulted by the finger, along with the other bright star (11.3 mag) in the field, South
of IN Com and close to the edge of the imaged field. Note the non-uniform distribution 
of nebular intensities. The nebulosity is much more extended 
West of the occulting finger than East of it. A similar asymmetric distribution of
intensities is apparent in the N-S direction, where the nebulosity is
more extended in the E-W direction, North of IN Com. The subjective asymmetry
is confirmed also by comparing ``typical'' cuts through the nebula; one in
the E-W direction across the Northern part of the nebula is brighter by a few
percent than one to the South of IN Com.

In general, if one could
reduce the nebulosity to its two principal components, it could be described as two
``blobs''. One lobe of the PN large and extends West of IN Com, North and 
South of the star. The other lobe is
much smaller and is located N-E of IN Com. There is sub-structure within the two 
blobs, but our images do not reveal this clearly. If anything, 
it is possible that the large blob to the West is composed of two smaller blobs. 
The structure of the
nebula is probably two-lobed, and possibly three-lobed. The same structure, 
by the way, is evident in the original image of LT80 but has not been
remarked on.  
The size of the nebulosity, as measured from our combined image, is 
$\sim$490"$\times$530". This is slightly smaller but rounder than given by LT80 
(490"$\times$560"). Note that it is difficult to mark exactly where the 
nebulosity ends and it is possible that we are missing the outer regions of the
PN, which are probably exceedingly faint.
 
In order to verify the reality of the asymmetric nebulosity we used a wavelet 
transform technique with filtering in wavelet space, to emphasize 
the large-scale features. Our image processing approach is to trim 
all the data to lie in between the 20 to 30 DNs, the range of the 
features seen when displaying different DN levels. 
 To suppress the shot noise of the image, a non-linear self-adaptive filter 
has been applied to the digital wavelet transform (DWT) of the
image using the Daubechies\_12 wavelet representation (Press \etal 
1992). The denoising algorithm is fully described in Hoffman (1998) 
and its various steps are illustrated in Fig. 2. This algorithm has 
recently been applied to the
analysis of the Sunyaev-Zeldovich effect and of the X-ray surface 
brightness of rich clusters of galaxies (Zaroubi {\it et al.} 1998).

The raw image is shown in the upper left panel of Fig. 2. The DWT is 
applied to the data and the absolute values of the wavelet 
coefficients are sorted in decreasing order. 
The sorted wavelet spectrum of the raw image is shown as
the solid line in the {\it log-log} plot shown in the lower
left panel of Fig. 2. Note the appearance of two approximate power 
laws, one for the $\approx$0.05\% highest amplitudes coeffcients, 
and a much shallower one for the rest of the coefficients. 
This is a typical behavior found in many astronomical 
and other kinds of images. Noisy images, characterized by distinct
non-random features, show a dual power law of the sorted wavelet spectrum. 
Clean, noiseless images of this kind exhibit $\sim$a single approximate 
power law behaviour of their sorted wavelet spectrum. 
The assumption made here, which is confirmed by the analysis of 
simulated images (Hoffman 1998), is that the high end of the
power-law corresponds to the clean input signal while the 
low end, with its shallower power law, is dominated by the noise. 

An adaptive  non-linear filter is applied to the 
wavelet coefficients such that the  low amplitude coefficients are supressed, 
in order to recover the single power law behaviour (Hoffman 1998).
This filter is non-linear, because it depends only on the amplitude of the 
coefficients and not on scale or position. It is self-adaptive, because 
it extrapolates the spectral high-end behaviour to the entire wavelet space, 
from approximately (5 10$^{-4}$, 20) in the lower left panel of Fig. 2.  
The horizontal line in the lower left panel of Fig. 2 shows the threshold 
value which distinguishes the signal-dominated from the noise-dominated 
regimes. The dashed line shows the sorted and filtered wavelet spectrum.
One should note the fundamental difference between the non-linear filter used
here and the commonly used linear filters. The latter operate by smoothing, 
namely suppressing some of Fourier (e.g.) modes according to their location
in the reciprocal space, thus leading to an inherent loss of resolution.
Non-linear filters, such as the one used here, suppress only the low amplitude
coefficients regardless of their scale and position, thus avoiding a loss of
resolution.

Fig. 2 shows the denoised image in the top right panel. A more 
quantitative presentation of the 
original and denoised images is given by the lower right panel, which 
shows horizontal and vertical line cuts passing near
the center of the denoised image. Two plots are presented, originating from
the noisy (dashed line) and denoised (solid line) image. One plot 
(corresponding to the horizontal cut) is arbitrary shifted in amplitude 
by 10 DNs relative to the other for the sake of clarity, and shows 
the amount of noise suppression achieved by the denoising algorithm.  
The H$\alpha$ image shows, after denoising, the basic structural features 
of the PN which were visible in our original image and in that of
LT80. The details are clearer in the denoised image, and the asymmetric light
distribution is easily percieved. Note that while the pattern of
``spillover'' light originating from the two stars hidden behind the
occulting finger is very similar, there are two brighter intensity
peaks in the West part of the PN, while on the East part only a
faint a minor peak is seen.
A more quantitive analysis would be  beyond the scope of this paper. 
 
Note in the top right panel of Fig. 2 that the low intensity patch East of 
IN Com is still present, although its shape has been somewhat modified by 
the properties of the transform. Some artefacts have been introduced, such as 
the two streaks parallel to the occulting finger and its extension to the
North, but in general the
image is easier to evaluate morphologically than the original one. Both the E-W
and N-S asymmetries have been retained in the wavelet-cleaned image; this
indicates that they are intrinsic to the nebulosity.  
 
\section{Discussion}

It is now possible to evaluate the PN morphology.
Adopting the  criteria set by Stanghellini \etal (1993), we note that
the images show (a) two main axes, one $\sim$N-S and the other
 $\sim$E-W,
and (b) a waist-like region near IN Com, oriented generally in the E-W direction.
Based on these, and considering the additional structure in the 
outer regions of the PN, one could classify LoTr 5 as a ``bipolar'' PN, perhaps
belonging to the subclass BM. In any case, the PN does not appear
to be symmetric.

 Asymmetry in a PN can be
caused by an interaction between the nebular gas and the ISM (Dgani \& Soker 1997).
The Atlas of Galactic Neutral Hydrogen (Hartmann \& Burton 1997) was consulted 
to establish whether the ISM in the immediate vicinity of LoTr 5 shows signs of 
disturbance. The only significant HI signal near IN Com appears at v$_{LSR}\approx0$ 
km s$^{-1}$ and the distribution of neutral hydrogen does not appear disturbed. 
For this reason, we 
believe that interaction with the ISM is not the shaping factor of this PN.
This was also the conclusion of Th\'{e}venin \& Kwitter (1996) based
on the LT80 image.

The morphology described above shows a peculiar ``hole'' East of IN Com, which
is visible in the LT80 photograph, in our original combined image, and in 
the wavelet reconstruction (though its shape changed slightly). 
In principle, a round feature could be a shadow cast onto the nebulosity by a 
foreground extended and opaque object, such as a dark globule. We consider this 
an unlikely  possibility, because both the IRAS Point Source Catalog and the Faint 
Source Catalog do not list a 100$\mu$m source at the high galactic latitude 
location of IN Com, and we believe this asymmetry to be intrinsic to the PN.
Soker (1997) reviewed different mechanisms which may shape PNs. 
A low surface brightness patch could be the result of a star within the nebula blowing 
a clear(er) patch through the PN material. This  is also  not a viable 
option for LoTr 5, as our images do not show a star within the ``hole''.
We also note that Soker classifed LoTr 5 as belonging to the common envelope
family; these objects are expected to form elliptical PNs, if the primary
survives to the AGB stage, and the hole-like feature could not be explained
in this case.
 
There are indications that the hot component of IN Com has a strong and fast stellar 
wind observed in high-excitation UV  lines (Modigliani \etal 1993; Feibelman 1994). 
The fast wind and the presence of high excitation
spectral lines indicate that this star is a few 10$^3$ years past its asymptotic giant
branch. It may be possible that the stellar wind played some role in the shaping of this
PN, although how it could have ``carved out'' the cavity it is not clear.

One of the latest analyses of the system (Jasniewicz \etal 1996) puts 
at the center of the PN a
close binary consisting of the hot sdO star and the G5III star, where the latter star spins
close to break-up velocity with P$_0$=5$^d$.9 and produces the photometric light
curve by its spotted surface. This raises another possibility of explaining the
shape on the PN through magnetic confinement of the stellar wind, which could be eased
in the case of a binary where the envelope was spun-up (cf. Livio 1997). 
The present-day sdO star could have been spun-up by tidal interaction with 
its companion in an earlier evolutionary phase, just as the companion
seems to be spun-up today. However, this possibility produces as a first 
approximation axisymmetrical configurations (Livio 1997),
not localized rarefactions in the PN as observed in this case.

\section{Conclusions}

A new image of the planetary nebula LoTr 5 (IN Com) was obtained in H$\alpha$ 
and [NII] light as a 
combination of CCD images taken with a coronagraph on the Wise Observatory telescope. 
The images shows an asymmetric nebula, arguing for the
presence of a close binary nucleus in this PN, with one component being the hot
star which ejected the nebula. Observations with EUVE will possibly establish the
temperature of the hot star and, through a comparison with models, its age on the 
cooling tracks. The mechanism which shaped this PN, and in particular created a
hole-like feature East of the central star, has not been identified.

\section*{Acknowledgements}
    UV research at Tel Aviv University is supported by grants from
    the Ministry of Science and Arts through the Israel Space Agency,
    from the Austrian Friends of Tel Aviv
    University, and from a Center of Excellence Award from the Israel
    Science Foundation. NB acknowledges support from a US-Israel Binational
    Award to study UV sources measured by the FAUST experiment, and the
hospitality of STScI during his sabbatical. 
YH acknowledges support from the US-Israel Binational Foundation (94-00185),
the  S.A. Schonbrunn Research Endowment Fund, and 
the Israel  Science Foundation (103/98). We acknowledge
the use of the CDS database at Strassbourg, in particular the on-line access to
the Hipparcos and Tycho catalogues. We are grateful to Mario Livio and Noam 
Soker for reading and commenting on an early draft of this paper. Steve Larson is
acknowledged for providing the original coronagraph, used for the 
observations presented here.

\newpage

\section*{References}
\begin{description}

\item Apparao, K.M.V., Berthiaume, G.D. \& Nousek, J.A. 1992, ApJ, 397, 534.

\item Bond, H.E. \& Livio, M. 1990, ApJ, 355, 568.

\item Brosch, N., Ofek, E.O., Almoznino, E., Sasseen, T.,  Lampton, M. \&
 Bowyer, S. 1998 MNRAS, 295, 959.

\item Buonatiro, L. 1993, A\&AS, 100, 531.

\item Cherepashchuk, A.M., Katisheva, N.A., Khruzina, T.S. \& Shugarov, S. Yu.
1996 ``Highly Evolved Close Binary Stars: Catalog'', Amsterdam: Gordon and Breach

\item Dgani, R. \& Soker, N. 1997, ApJ, 484, 277.

\item Feibelman, W.A. 1994, PASP, 106, 56.

\item Feibelman, W.A. \& Kaler, J.B. 1983, ApJ, 269, 592.

\item Fruscione, A., Drake, J.J., McDonald, K. \& Malina, R.F. 1995,
ApJ, 441, 726.

\item Hartmann, D. \& Burton, W.B. 1997 ``Atlas of Galactic Neutral
Hydrogen'', Cambridge: Cambridge University Press.

\item Hoffman, Y. 1998, preprint.

\item Jasniewicz, G., Duquennoy, A. \& Acker, A. 1987, A\&A, 180, 145.

\item Jasniewicz, G., Acker, A., Mauron, N., Duquennoy, A. \& Cuypers, J.
1994, A\&A, 286, 211.

\item Jasniewicz, G., Th\'{e}venin, F., Monier, R. \& Skiff, B.A. 1996,
A\&A, 307, 200.

\item Kreysing, H.C., Diesch, C., Zweigle, J. \etal 1992, A\&A, 264, 623.

\item Livio, M. 1997 Space Sci. Rev., 82, 389.

\item Longmore, A.J. \& Tritton, S.B. 1980, MNRAS, 193, 521.

\item Malasan, H.L., Yamasaki, A. \& Kondo, M. 1991, AJ, 101, 2131.

\item Modigliani, A., Patriarchi, P. \& Perinotto, M. 1993, ApJ, 415, 258.

\item Press, W.H., Teukolsky, S.A., Vetterling, W.T.
\& Flannery, B.P. 1992, ``Numerical Recipes in FORTRAN'', Cambridge: Cambridge
University Press (2nd edition).

\item Schnell, A.M. \& Purgathofer, A. 1983, A\&A, 127, L5.

\item Soker, N. 1997, ApJS, 112, 487.

\item Stanghellini, L., Corradi, R.L.M. \& Schwarz, H.E. 1993, A\&A, 279, 521.

\item Strassmeier, K.G., Bartus, J., Cutispoto, G. \& Rodon\'{o}, M.
1997a, A\&AS, 125, 11.

\item Strassmeier, K.G., Hubl, B. \& Rice, J.B. 1997b, A\&A, 322, 511.

\item Th\'{e}venin, F. \& Jasniewicz, G. 1997, A\&A, 320, 912.

\item Tweedy, R.W. \& Kwitter, K.B. 1996, ApJS, 107, 255,

\item Zarubi, S., Sqyires, G., Hoffman, Y. \& Silk, J. 1998, ApJL, in press.


\end{description}

\newpage

\section*{Figure captions}
\begin{itemize}
\item Figure 1: H$\alpha$ surface brightness image of IN Com and the planetary 
nebula LoTr 5. East is 
up and North is to the right. The diameter of the circular image is 10.5 arcmin,
and this is the only illuminated part of the square CCD. IN Com
is the bright object close to the center of the field and its light is mostly blocked 
by the occulting finger.
 
\item Figure 2: H$\alpha$ image of IN Com and  LoTr 5; original
(top left) and after wavelet cleaning (top right). The image 
parameters are as for Fig. 1. The lower panels show the wavelet spectrum
(solid line, lower left) and its truncation with extension 
(dashed line), and two cuts through the center of the image (North-South=upper 
and East-West=lower), with the original data as the dashed 
jiggly line and the denoised one as the heavy line. The cuts have been 
shifted vertically by 10 DNs to separate them clearly on the plot.

\end{itemize}

\newpage


\begin{figure}[tbh]
\vspace{11cm}
\includegraphics{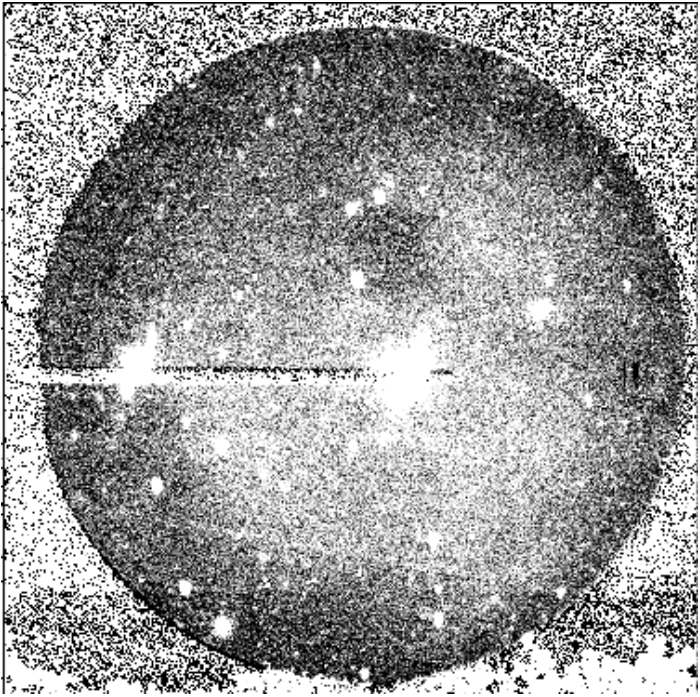}
\caption{\it {H$\alpha$ surface brightness image of IN Com and the planetary 
nebula LoTr 5. East is 
up and North is to the right. The diameter of the circular image is 10.5 arcmin,
and this is the only illuminated part of the square CCD. IN Com
is the bright object close to the center of the field and its light is mostly blocked 
by the occulting finger.}}
\end{figure}

\newpage

\begin{figure}[htb]
\putplot{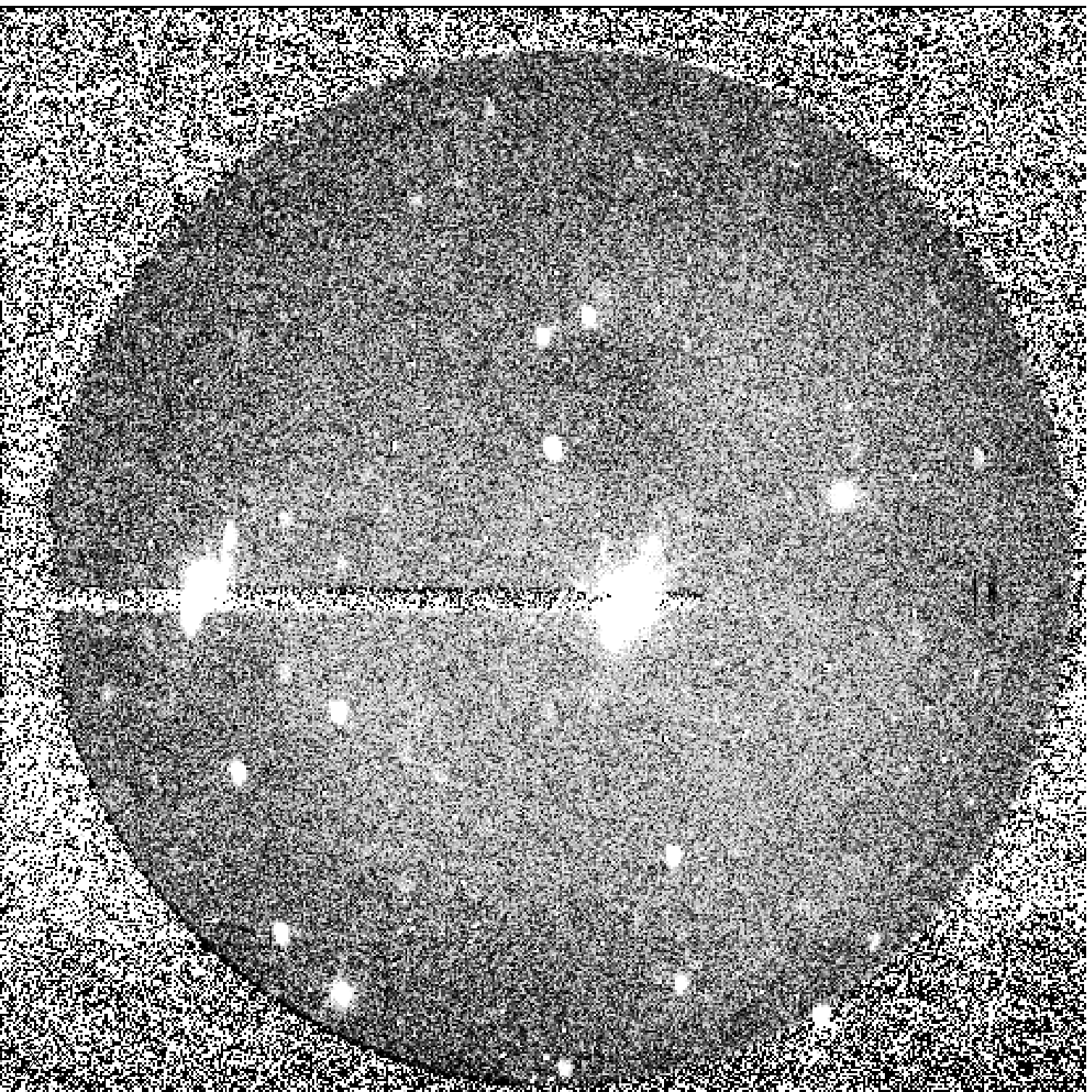}{1in}{0}{50}{50}{-250}{-130}
\putplot{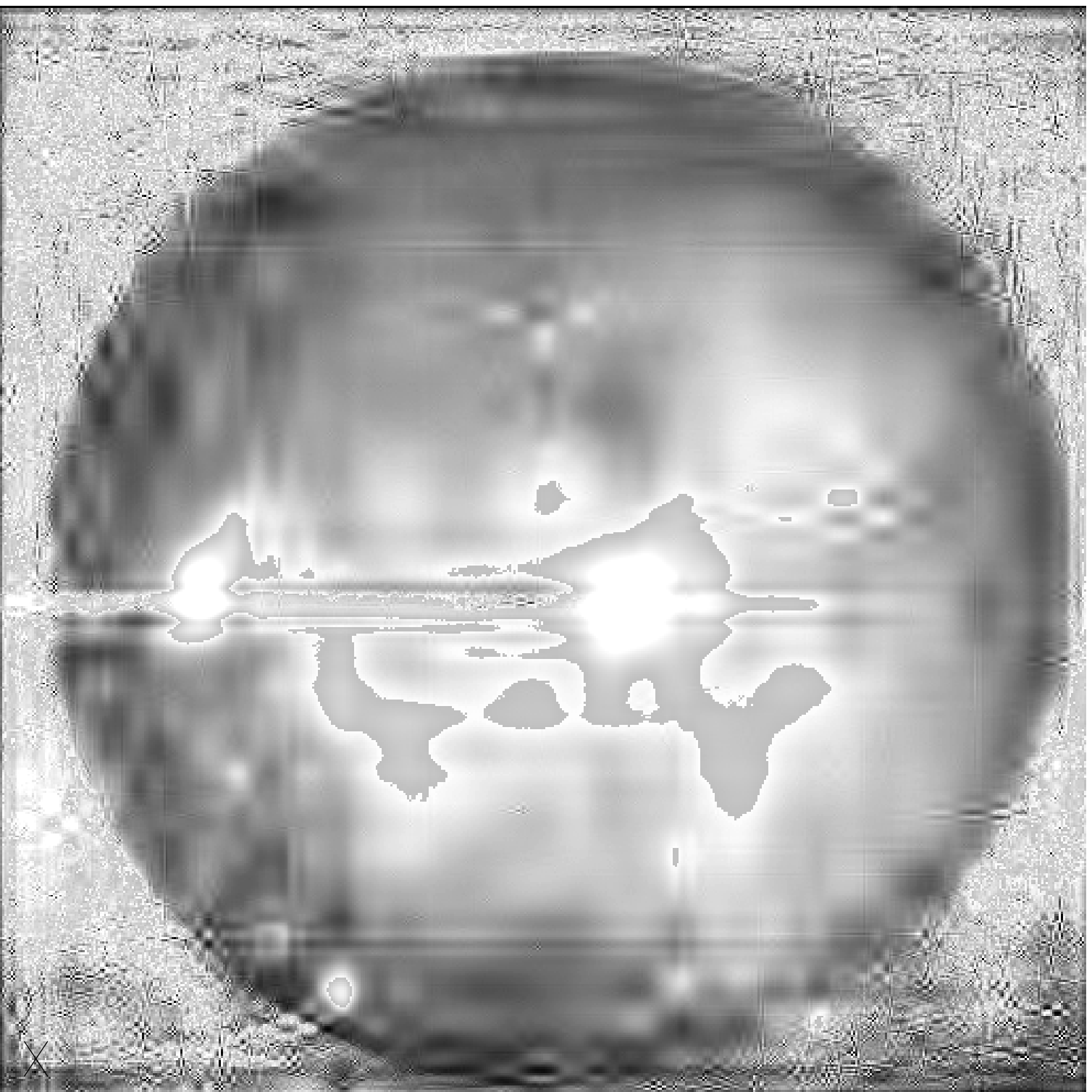}{1in}{0}{50}{50}{0}{-37}
\putplot{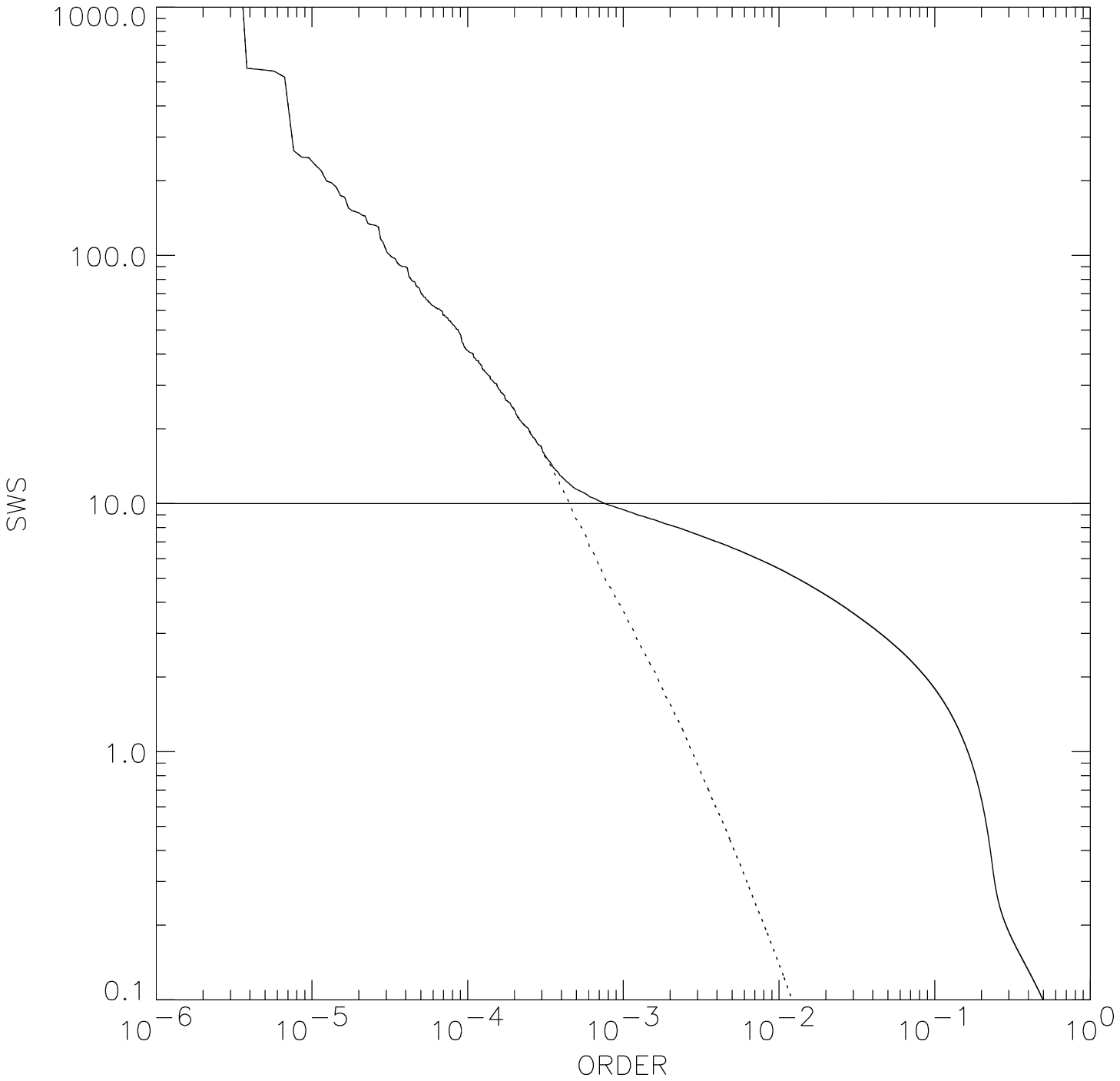}{1in}{0}{50}{50}{-250}{-230}
\putplot{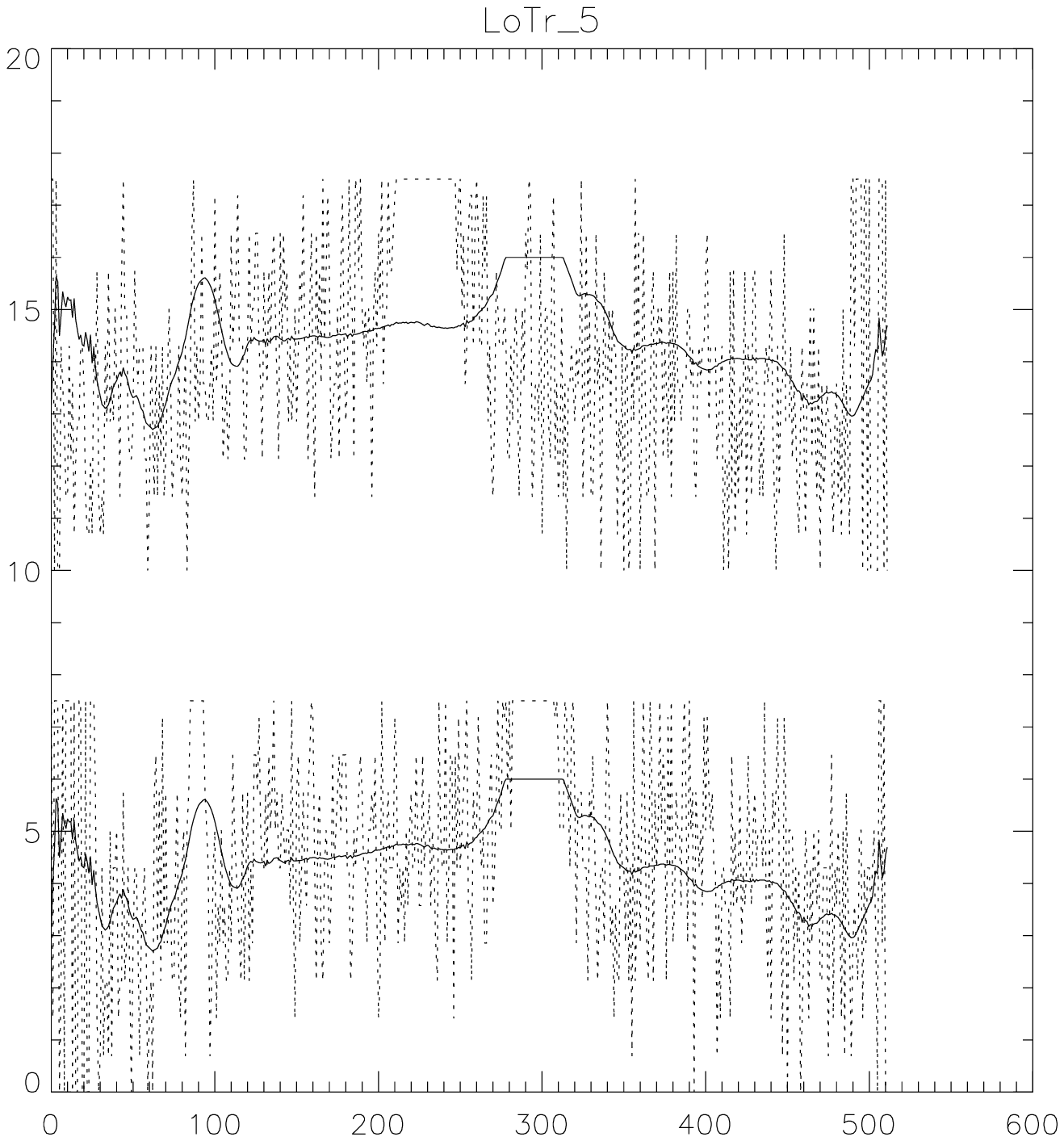}{1in}{0}{50}{50}{0}{-137}

\end{figure}

 \newpage

\end{document}